\documentclass[twocolumn,aps,prc,showpacs,floatfix]{revtex4}
\usepackage{amsmath,bm}
\usepackage{graphicx}
\usepackage{epsfig}
\begin{document}

\title{Role of isospin physics in supernova matter and neutron stars}
\author{Bharat K. Sharma and Subrata Pal}
\affiliation{Department of Nuclear and Atomic Physics, Tata Institute of 
Fundamental Research, Homi Bhabha Road, Mumbai 400005, India}

\begin{abstract}
We investigate the liquid-gas phase transition of hot protoneutron stars shortly after 
their birth following supernova explosion and the composition and structure of hyperon-rich
(proto)neutron stars within a relativistic mean-field model where the nuclear symmetry energy 
has been constrained from the measured neutron skin thickness of finite nuclei. Light clusters 
are abundantly formed with increasing temperature well inside the neutrino-sphere 
for an uniform supernova matter. Liquid-gas phase transition is found to suppress the cluster 
yield within the coexistence phase as well as decrease considerably the neutron-proton 
asymmetry over a wide density range. We find symmetry energy has a modest effect on the 
boundaries and the critical temperature for the liquid-gas phase transition, and the composition 
depends more sensitively on the number of trapped neutrinos and temperature of the protoneutron 
star. The influence of hyperons in the dense interior of stars makes the overall equation
of state soft. However, neutrino trapping distinctly delays the appearance of hyperons
due to abundance of electrons. We also find that a softer symmetry energy further 
makes the onset of hyperon less favorable. The resulting structures of the (proto)neutron
stars with hyperons and with liquid-gas phase transition are discussed.

\end{abstract}

\maketitle

\section{Introduction}

Recently considerable effort has been devoted to understand the isospin
physics because of its paramount importance in determining the structure and 
dynamics of finite nuclei \cite{Danielewicz,Baran,BALi08} and also in the 
astrophysical contexts such as supernova dynamics, protoneutron star evolutions 
and the structure and properties of neutron star \cite{Glend,Prakash,Steiner,Lattimer}. 
In stable nuclei, the neutron-proton asymmetry, $\delta = (\rho_n-\rho_p)/(\rho_n+\rho_p)$, 
is about 0.2, where $\rho_n$ and $\rho_p$ are the neutron and proton number densities.
In forthcoming rare-isotope accelerator experiments the magnitude of $\delta$ could be 
well above 0.24. In contrast, a large $\delta \approx 0.9$ is reached in the dense interior 
of neutron stars \cite{Glend,Steiner,Schaffner}. Whereas, very small asymmetry of 
$\delta \sim 0.1$ is attained near the crust of (proto)neutron stars 
at densities $\rho \sim 10^{-5}$-$10^{-1}$ fm$^{-3}$ \cite{Burrows,Keil}. 
In order to understand the composition and structure of stars over
this extreme values of  density and asymmetry requires precise knowledge of
the density dependence of nuclear symmetry energy $E_{\rm sym}(\rho)$.

Several theoretical studies have demonstrated \cite{Baran,BALi08} that
$E_{\rm sym}(\rho)$ represents the leading coefficient of an expansion of the
total energy, $E(\rho,\delta$), for a neutron-rich system with respect to
asymmetry: $E(\rho,\delta) = E(\rho,0) + E_{\rm sym}(\rho)\delta^2 + 
{\mathcal{O}}(\delta^4)$. The symmetric nuclear matter equation of state, $E(\rho,0)$,
has been already well constrained over a wide density range from analysis of a large 
body of data within non-relativistic models \cite{Blaizot,PrakBed}
and relativistic models \cite{Danielewicz}. In contrast, though the nuclear symmetry
energy at the normal nuclear matter density $\rho_0=0.15$ fm$^{-3}$ is well 
constrained to $E_{\rm sym}(\rho_0)=32\pm 4$ MeV,
its value and trend at lower and especially at higher densities are poorly known 
\cite{Brown,Danielewicz}. Various theoretical calculations based on microscopic
\cite{Bruck,Siemens,Akmal} and/or phenomenological many-body approaches
\cite{Baran,BALi08,Brown,Serot,Pak,BALi97}, though reproduce the well-constrained 
value of $E_{\rm sym}(\rho_0)$, their high density predictions of symmetry energy
are extremely diverse. While certain models predict a monotonically increasing 
symmetry energy with density, other model results show that $E_{\rm sym}(\rho)$ 
increases initially up to about $\rho_0$ and decreases thereafter. 
These diverse predictions in the density dependence in the symmetry energy lead 
to large uncertainty in the incompressibility of symmetric nuclear matter
which mainly stems due to insufficient knowledge about the isospin dependence
of in-medium nuclear effective interactions.

Recently some progress has been achieved by consistently constraining the symmetry
energy to $32(\rho/\rho_0)^{0.7} \leq E_{\rm sym}(\rho) \leq 32(\rho/\rho_0)^{1.1}$
at subsaturation densities from analysis of isospin diffusion \cite{BALi08,Chen}
and isoscaling \cite{Shetty} data in intermediate energy heavy ion collisions
and from the study of neutron skin of several nuclei \cite{Cent,SharmaPLB}. 
Within a relativistic mean-field (RMF) model \cite{Serot,NL3,FSU}, the symmetry energy 
has been constrained to a small range \cite{SharmaPLB} from analysis of neutron skin 
thickness of several nuclei across the periodic table. Further, the slope parameter
and the incompressibility for the constrained $E_{\rm sym}(\rho)$ was found 
\cite{SharmaPLB} consistent with that extracted from isospin diffusion and isoscaling data.

In an effort to pindown the precise density dependence of symmetry energy, it has been
noted \cite{Brown,Horowitz} that the skin thickness of neutron-rich heavy nucleus, 
for example $^{208}$Pb, are linearly correlated with the symmetry pressure at a 
density slightly below the normal nuclear matter value. Since the same pressure
supports a neutron star at supranormal densities against gravitational collapse,
suggests an underlying correlation between neutron skin of heavy nuclei and 
neutron star radii \cite{Horowitz}. Within the RMF models, the constrained 
$E_{\rm sym}(\rho)$ demonstrated \cite{SharmaPLB} that the neutron skin in $^{208}$Pb
tend to yield smaller neutron star radii. Furthermore, within the constrained range
of symmetry energy, significant sensitivity on several features of liquid-gas phase 
transition \cite{Mueller,Wang} in hot asymmetric nuclear matter was found 
\cite{SharmaPRC}. In particular, the boundary and area of the liquid-gas coexistence 
region, the maximal isospin asymmetry and the critical values of pressure and 
isospin asymmetry all of which systematically increase with increasing overall 
softness of $E_{\rm sym}(\rho)$.

It is thus instructive to explore the effects of symmetry
energy in the RMF model, that has been limited to a narrow range from 
analysis of skin data, on the composition and structure of hot protoneutron
and cold neutron star matter that encompasses a large density range. 
A newly born neutron star or protoneutron star (PNS) 
\cite{Prakash,Burrows,Keil,Pons} is formed in the aftermath
of a successful supernova explosion and subsequent gravitational collapse
of the core of a massive star. Due to short mean free path of neutrinos
of $\lambda_\nu \sim 10$ cm  (compared to PNS radii of about 10 km), the
neutrinos are trapped temporarily within the neutrino-sphere of the star.
The PNS are thus rich in leptons, mostly $e^-$ and $\nu_e$, and quite hot 
with central temperatures of $T = 10-50$ MeV. Thus the star matter has the 
interesting possibility to undergo liquid-gas phase transition close to the
surface of the star \cite{Ishi,Botvina}, the characteristics of which is
mainly controlled by the symmetry energy at subsaturation densities.
In the much longer Kelvin-Helmholtz evolution stage of about 10 sec,
the neutrinos escape from the interior, and the hot and lepton-rich
PNS changes into a cold and deleptonized neutron star. The loss of neutrinos
enforce the electrons and protons to combine resulting in a neutron-rich
neutron star matter. Consequently, the symmetry energy at supranormal
densities plays an important role in determining the composition,
in particular the possible presence of strange hyperons, the structure
(mass and radius) and the long-term cooling of the neutron star. 

In this paper we study the role of isospin on the liquid-gas phase transition
in hot lepton-rich protoneutron star and the properties of (proto)neutron star
within the accurately calibrated parameter set FSUGold \cite{FSU} in the relativistic 
mean-field (RMF) model. For this purpose, we restrict the model density dependence 
of symmetry energy $E_{\rm sym}(\rho)$ within a narrow range that reproduced the skin
thickness of several nuclei \cite{SharmaPLB}.
At the supranormal densities the contribution of the hyperon degrees of freedom 
to the equation of state of the star is considered. At very low densities well below
the saturation value, that is relevant to the crust, few-body correlations become 
important. The system can then minimize its energy via the formation of
medium-modified light clusters \cite{Ropke,Typel} which eventually dissolve into 
homogeneous nuclear matter at higher densities due to Pauli blocking.

The paper is organized as follows:
In Sec. II we introduce the RMF model for baryons along with the medium-modified
clusters. The construction of the liquid-gas phase transition in the hot 
protoneutron star matter is presented. 
In Sec. III the numerical results for the structure and composition
of the star matter are discussed. Sec. IV summarizes the results. 
\bigskip

\section{Formalism}
\subsection{The RMF Model with clusters} 

We consider the relativistic mean-field (RMF) model with all the charge states of the baryon 
octet ($B \equiv p,n,\Lambda,\Sigma^+,\Sigma^0,\Sigma^-,\Xi^0,\Xi^-$) and their 
antiparticles. Light clusters ($C \equiv ~^2{\rm H}, ~^3{\rm H}, ~^3{\rm He}, 
~^4{\rm He} \equiv d,t,h,\alpha$), that are treated as 
quasi-particles, are also included as explicit degrees of freedom. The properties 
of these clusters are modified by medium effects via the temperature-dependent 
shifts in the binding energies as treated in Ref. \cite{Ropke,Typel}. 
The interaction Lagrangian density in the nonlinear RMF model 
\cite{FSU,Typel,Fatt} for a system of baryons, clusters and leptons for protoneutron 
star matter is given by 
\begin{widetext}
\begin{eqnarray} \label{RMF}
\mathcal{L} &=& \sum_{i=B,t,h} {\overline\psi}_i 
\left( \gamma_\mu iD_i^\mu - m^*_i \right) \psi_i 
+ \frac{1}{2} (iD_\alpha^\mu\phi_\alpha)^* (iD_{\alpha\mu}\phi_\alpha) 
- \frac{1}{2} m_\alpha^{*2} \phi_\alpha^* \phi_\alpha 
+ \frac{1}{4} (iD_d^\mu\phi_d^\nu - iD_d^\nu\phi_d^\mu)^* 
(iD_{d\mu}\phi_{d\nu} - iD_{d\nu}\phi_{d\mu}) \nonumber\\ 
&& - \frac{1}{2} m_d^{*2} \phi_d^{\mu*} \phi_{d\mu}
+ \frac{1}{2} ( \partial_\mu \sigma \partial^\mu \sigma - m_\sigma^2\sigma^2 )
-\frac{\kappa}{3!}\left(g_{\sigma N} \sigma\right)^3
- \frac{\lambda}{4!}\left(g_{\sigma N} \sigma\right)^4 
- \frac{1}{4} {\omega}_{\mu\nu} {\omega}^{\mu\nu} 
+ \frac{1}{2} m_\omega^2 {\omega}_\mu {\omega}^\mu  
+ \frac{\zeta}{4!} \left(g_{\omega N}^2 \omega_\mu \omega^\mu \right)^2 \nonumber \\
&& - \frac{1}{4} {\bm\rho}_{\mu\nu} \cdot {\bm\rho}^{\mu\nu} 
+ \frac{1}{2} m_\rho^2 {\bm\rho}_\mu \cdot {\bm\rho}^\mu   
+ \Lambda_v \left( g_{\rho N}^2 {\bm\rho}_\mu \cdot {\bm\rho}^\mu \right)
\left(g_{\omega N}^2 \omega_\mu \omega^\mu \right) 
~ + \sum_l {\overline\psi}_l \Big( i \gamma_\mu \partial^\mu - m_l \Big) \psi_l ,
\end{eqnarray}
\end{widetext}
where $\psi_i$ is spin-1/2 fields for the particles ($i=B,t,h$), $\phi_\alpha$
is the spin-0 field for the alpha particle, and $\phi_d^\mu$ is the spin-1
field for deuteron all of which are coupled to the $\sigma,\omega,\rho$ mesons. 
The field strength tensors for the vector mesons are 
${\omega}_{\mu\nu} = \partial_\mu\omega_\nu - \partial_\nu\omega_\mu$ and 
${\bm\rho}_{\mu\nu} = \partial_\mu {\bm \rho}_\nu - \partial_\nu {\bm\rho}_\mu$. 
The nonlinear $\sigma$-meson couplings ($\kappa, \lambda$) soften the symmetric 
nuclear matter equation of state (EOS) at around $\rho_0$, while its high density 
part is softened by the self-interactions ($\zeta$) for the $\omega$-meson field. 
The sum on $l$ is over the non-interacting electrons and muons ($e^-$ and $\mu^-$) 
and their antiparticles in the star. 

The covariant derivative
\begin{equation} \label{covd}
iD^\mu_i = i\partial^\mu - g_{\omega i}A_i\omega^\mu 
- g_{\rho i} |N_i-Z_i| {\bm\tau} \! \cdot \! {\bm\rho}^\mu I_{3i},
\end{equation}
of the particle $i$ with mass, proton, neutron numbers ($A_i,Z_i,N_i$)
involves interaction with the vector mesons. Of course for hyperons
($i\equiv Y$) the ($A_i,Z_i,N_i$) drop out in Eq. (\ref{covd}) and hereafter.
$I_{3i}$ is the isospin projection of the baryon and cluster charge state $i$. 
The effective mass of the $i$th species is 
\begin{equation} \label{effm}
m_i^* = m_i - g_{\sigma i}A_i\sigma - \Delta B_i ,
\end{equation}
where the vacuum rest mass of the clusters $i=d,t,h,\alpha$ is given
by $m_i = Z_im_p + N_im_n - B_i^0$. For the vacuum binding energy $B_i^0$
of the clusters we adopt the experimentally measured values. Note the clusters 
which are predominant at densities $\rho \sim (10^{-5}-10^{-1})\rho_0$ are treated 
as structureless point particles. The scalar isovector $a_0(980)$ (the $\delta$ meson) 
that splits the effective masses of the baryons has not been included in the present study
as its effects on the star properties were found to be minimal \cite{Schaffner,Pais}.
The medium dependent shift of the binding energy $\Delta B_i$ that appears only for 
the clusters are adopted from Ref. \cite{Typel}. For nuclei embedded in nuclear matter,
the shift has been calculated by solving the in-medium Schr\"odinger equation 
which leads to the ground state energy \cite{Typel}
\begin{equation} \label{shift}
B_{A,Z} = B_{A,Z}^0 + \frac{K^2}{2Am} + \Delta B^{SE}_{A,Z} + \Delta B^{Pau}_{A,Z} 
+ \Delta B^{Cou}_{A,Z}. 
\end{equation}
The shift due to self-energy, $\Delta B^{SE}_{A,Z}$, is already contained in the effective 
particle mass in the RMF model. The Coulomb shift, $\Delta B^{Cou}_{A,Z}$, that can be 
evaluated within the Wigner-Seitz approximation gives a small value for the light clusters 
and thus neglected. The dominant contribution emerge from the Pauli shift,
$\Delta B^{Pau}_{A,Z}$, that has been evaluated in the perturbation theory 
with Jastrow and Gaussian approaches for $d$ and ($t,h,\alpha$), respectively.
We adopt the empirical form for Ref. \cite{Typel} 
\begin{equation} \label{deltb}
\Delta B_i(\rho_p^{\rm tot}, \rho_n^{\rm tot}, T) = - \tilde{\rho}_i 
\Big[1 + \frac{\tilde{\rho}_i}{2\tilde{\rho}_i^0(T)}\Big] \delta B_i(T) ,
\end{equation}
where $\tilde{\rho}_i = 2(Z_i\rho_p^{\rm tot} + N_i\rho_n^{\rm tot})/A_i$ and 
$\tilde{\rho}_i^0(T) = B_i^0/\delta B_i(T)$. The quantity 
$\delta B_i(T)$ can be found from Eqs. (26)-(27) in Ref. \cite{Typel}. With
increasing temperature and/or density a cluster will dissolve when its
total binding energy $B_i = B_i^0 + \Delta B_i$ vanishes.
In the above expression the total densities for neutrons, 
$\rho_n^{\rm tot} = \sum_{i=B,C} N_i\rho_i$, and protons, 
$\rho_p^{\rm tot} = \sum_{i=B,C} Z_i\rho_i$ are contributions from baryons ($B$)
and clusters ($C$).

The composition of the star is determined by the requirements of charge neutrality and 
$\beta$-equilibrium conditions $B_1 \to B_2 + l + {\overline\nu}_l$, 
$B_2 + l \to B_1 + \nu_l$, where $B_1$ and $B_2$ are the baryons and $l$
is a lepton ($e^-$ or $\mu^-$) or their respective antiparticles at finite 
temperature. With clusters ($A,Z$), these weak processes correspond to
$(A,Z) \to (A,Z+1) + l + {\overline\nu}_l$ and  
$(A,Z) + l \to (A,Z-1) + \nu_l$. The chemical potential of a species
$i$ with baryon number $B_i$, charge $Q_i$, and lepton number $L_i$ in chemical
equilibrium is in general  
\begin{equation} \label{gchem}
\mu_i = B_i\mu_B + Q_i\mu_Q + L_i \mu_L .
\end{equation}
The three independent chemical potentials $\mu_B$, $\mu_Q$ and $\mu_L$ can be determined 
from total baryon number, charge, and lepton number conservations leading to
\begin{eqnarray} \label{chem}
&&\mu_{A,Z} =  A\mu_B + Z \mu_Q ,  \nonumber\\
&&\mu_{e^-} = - \mu_{e^+}  = -\mu_Q + \mu_L ,  \nonumber\\
&&\mu_{\nu_l} = - \mu_{{\overline\nu}_l}  = \mu_L ; ~~~ (l=e^-,\mu^-) .
\end{eqnarray} 
For a deleptonized neutron star, the (anti)neutrinos escape freely so that
$\mu_L = 0$. The remaining chemical potentials, $\mu_B$ and $\mu_Q$, can be 
obtained from conservations of total baryon and charge densities
\begin{eqnarray} \label{consv}
&& \rho = \sum_{i=B,C} A_i \rho_i = \rho_n^{\rm tot} + \rho_p^{\rm tot} , \nonumber\\
&& \rho_Q = \sum_{i=B,C} Q_i \rho_i  + \sum_{l=e,\mu} Q_l \rho_l  = 0 . 
\end{eqnarray} 

When neutrinos are trapped, the matter (inside the neutrino-sphere) 
on a dynamical time scale is characterized by a fixed lepton fraction
(number of leptons per baryon), $Y_{Ll} = Y_L + Y_{\nu_l}$, for each flavor
$l=e,\mu$. In fact, muons are absent in a PNS that leads to $Y_{L\mu} = 0$ 
and $\mu_{\nu_l} \equiv \mu_{\nu_e}$ in Eq. (\ref{chem}).
Consequently, the composition of a hot PNS
is characterized by a constant value of $Y_{Le} = Y_e + Y_{\nu_e}$, which
in the late stage reaches a value of $Y_{Le} \simeq 0.4$.

The number density for a fermionic species $i=B,t,h,l$ at temperature $T\neq 0$
\begin{equation} \label{fdens}
\rho_i = \frac{2J_i+1}{(2\pi)^3} \int d^3k \: \left[ f_i^+(k) - f_i^-(k) \right] ,
\end{equation}
where $J_i$ is the spin of the $i$th species and the distribution function 
for particles and antiparticles (referred to as $\pm$ sign) is as usual 
\cite{Mueller,Wang}
\begin{equation} \label{DF}
f_i^\pm(k) = \left[ \exp\left\{ \left(E^*_i(k)\mp \nu_i\right)/T\right\}+1 \right]^{-1}.
\end{equation}
The density of the bosonic species $i=d,\alpha$ is obtained from \cite{Typel}
\begin{equation} \label{bdens}
\rho_i = \frac{2J_i+1}{(2\pi)^3} \int d^3k \: g_i(k) + \tilde{\rho}_i ,
\end{equation}
where $\tilde{\rho}_i$ stems from the condensed bosons in the ground state.
The Bose-Einstein distribution follows as
\begin{equation} \label{DB}
g_i(k) = \left[ \exp\left\{ \left(E^*_i(k) - \nu_i\right)/T \right\} - 1 \right]^{-1} .
\end{equation}
For interacting baryons and clusters, the effective energy and chemical potentials are 
respectively $E^*_i = \sqrt{k^2+m_i^{* 2}}$ and 
$\nu_i = \mu_{A,Z} - g_{\omega i}A_i\omega_0 - g_{\rho i}(N_i-Z_i)\rho_{03}I_{3i}$, where
$\omega_0$ and $\rho_{03}$ are the timelike and isospin three component of
the vector $\omega$-meson and vector-isovector $\rho$-meson. For field-free lepton, these 
quantities reduce to $E^*_l \equiv E_l = \sqrt{k^2+m_l^2}$ and $\nu_l = \mu_l$. 
The field equations for the mesons ($m=\sigma,\omega, \rho$) are derived in the usual 
way \cite{Mueller} for the Lagrangian of Eq. (\ref{RMF}). Herein the contribution from the 
cluster densities to the source terms in the meson fields are consistently incorporated 
by noting that in the present Lagrangian the couplings $g_{mi}$ of the meson 
fields to the baryons and clusters ($i=B,d,t,h,\alpha$) are constant (density independent) 
unlike that in Ref. \cite{Typel}.
 
At finite temperature and baryon density, the energy density and pressure corresponding to 
the Lagrangian of Eq. (\ref{RMF}) can be obtained from the thermodynamical potential 
$\Omega$ \cite{Mueller} as
\begin{eqnarray} \label{eden}
{\cal E} & = & 
\frac{m_\sigma^2 \sigma^2}{2} + \frac{\kappa}{3!} \left(g_{\sigma N}\sigma\right)^3
+ \frac{\lambda}{4!} \left(g_{\sigma N} \sigma\right)^4  
+ \frac{m_\omega^2 \omega_0^2}{2} \nonumber\\
&& + \frac{\zeta}{8}\left(g_{\omega N} \omega_0\right)^4 
+ \frac{m_\rho^2 \rho_{03}^2}{2} 
+ 3\Lambda_v\left( g_{\omega N} \omega_0\right)^2 \left(g_{\rho N}\rho_{03} \right)^2
\nonumber\\
&& + \sum_{i=B,t,h,l} \frac{2J_i+1}{(2\pi)^3} \int d^3k \: E^*_i(k) 
\left[ f_i^+(k) + f_i^-(k) \right]  \nonumber\\
&& + \sum_{i=d,\alpha} \left[ \frac{2J_i+1}{(2\pi)^3} \int d^3k \: E^*_i(k) g_i(k) 
+ m^*_i\tilde{\rho}_i \right] ,
\end{eqnarray}
\begin{eqnarray} \label{pden}
 P & = & 
- \frac{m_\sigma^2 \sigma^2}{2} - \frac{\kappa}{3!} \left(g_{\sigma N}\sigma\right)^3
- \frac{\lambda}{4!} \left(g_{\sigma N} \sigma\right)^4  
+ \frac{m_\omega^2 \omega_0^2}{2} \nonumber\\
&& + \frac{\zeta}{24}\left(g_{\omega N} \omega_0\right)^4 
+ \frac{m_\rho^2 \rho_{03}^2}{2} 
+ \Lambda_v\left( g_{\omega N} \omega_0\right)^2 \left(g_{\rho N}\rho_{03} \right)^2
\nonumber\\
&& + \sum_{i=B,t,h,l} \frac{2J_i+1}{3(2\pi)^3} \int d^3k \: \frac{k^2}{E^*_i(k)} 
\left[ f_i^+(k)  + f_i^-(k) \right] \nonumber\\
&& + \sum_{i=d,\alpha} \frac{2J_i+1}{3(2\pi)^3} \int d^3k \: \frac{k^2}{E^*_i(k)} g_i(k) .
\end{eqnarray}

\subsection{Liquid-gas phase transition in PNS matter}

The RMF model is applied to study the liquid-phase (LGP) transition in hot
protoneutron star matter. Such a coexistence phase may occur at subsaturation densities 
where the star is composed of nucleons ($N$), clusters ($C$) and leptons only. A single-phase
system at temperature $T$ and density $\rho$ is stable against LGP separation
if its free energy density $\cal F$ is lower than the coexistence liquid ($L$)
and gas ($G$) phase:
\begin{equation} \label{free}
{\cal F}(T,\rho,\rho_3) < (1-\chi) {\cal F}^L(T,\rho^L,\rho_3^L) 
+ \chi {\cal F}^G(T,\rho^G,\rho_3^G) .
\end{equation} 
The total baryon density, $\rho=\rho_p^{\rm tot}+\rho_n^{\rm tot}$, and the isospin density, 
$\rho_3=\rho_p^{\rm tot} -\rho_n^{\rm tot} = -\rho\delta$, for a $N$-$C$-$e$-$\nu_e$ 
system satisfy
\begin{eqnarray} \label{den}
\rho &=& (1-\chi) \rho^L + \chi \rho^G \nonumber\\[1mm]
\rho_3 &=& (1-\chi) \rho_3^L + \chi \rho_3^G ,
\end{eqnarray}
where $\chi = V^G/V$ being the fraction of the total volume occupied by 
the gas phase. The stability condition, Eq. (\ref{free}), implies the inequalities 
\cite{Mueller} $\partial^2 {\cal F}/\partial\rho^2 > 0$ and 
$(\partial^2{\cal F}/\partial\rho^2) (\partial^2{\cal F}/\partial\rho_3^2)  > 
(\partial^2{\cal F}/\partial\rho_3 \partial\rho)^2 $. These inequalities
are equivalent to 
\begin{eqnarray} 
\rho\left(\frac{\partial P}{\partial \rho}\right)_{T,\delta}
&=& \rho^2 \left( \frac{\partial^2{\cal F}}{\partial\rho^2} \right)_{T,\delta} > 0 , 
\label{MEC} \\[1mm]
\left(\frac{\partial \mu_p}{\partial \delta}\right)_{T,P} < 0
~ &{\rm or}& ~
\left(\frac{\partial \mu_n}{\partial \delta}\right)_{T,P} > 0 .  \label{CHM}
\end{eqnarray}
The first inequality indicates mechanical stability which means a system
at positive isothermal compressibility remains stable at all densities. The second
inequality corresponds to chemical stability which suggests that energy is required
to alter the concentration in a stable system while maintaining temperature and pressure
fixed. When one of these conditions get violated, a two-phase system is
energetically favorable. The coexistence liquid-gas phase is governed by the Gibbs 
criteria for equal pressures and chemical potentials in the two phases with 
different densities but at the same temperature:
\begin{eqnarray} 
P(T, \rho^L) &=& P(T,\rho^G) ,  \label{GBS1} \\[1mm]
\mu_i(T,\rho^L) &=& \mu_i(T,\rho^G) , \label{GBS2} 
\end{eqnarray} 
where the chemical equilibrium condition refers to all the species 
($i = n,p,d,t,h,\alpha,e,\nu_e$) in the system.
Since we are dealing with free leptons, it is evident from Eq. (\ref{GBS2})
that the lepton density in the liquid and gas phases are identical
and thus forms a common background in the baryonic matter. Else for interacting leptons,
$\rho_l^L \neq \rho_l^G$, and thus the lepton number conservation would result 
in an additional condition similar to those of Eq. (\ref{den}).

\subsection{Symmetry energy and model parameters} 

The original FSUGold \cite{FSU}, with an isovector coupling $\Lambda_v=0.03$, 
produces a soft symmetric nuclear EOS with an incompressibility $K_0=230$ MeV.
In addition to the ground state properties of several nuclei, this model reproduces 
the strengths of the giant monopole resonance in $^{90}$Zr and $^{208}$Pb.
The nucleonic matter symmetry energy for this Lagrangian at temperature $T=0$ is given by
\begin{equation} \label{esym}
E_{\rm sym}(\rho) = \frac{k_{FN}^2}{6E_{FN}^*} + 
\frac{g_{\rho N}^2}{12\pi^2} \frac{k_{FN}^3}{m_\rho^{*2}} ,
\end{equation}
where $E_{FN}^* = \sqrt{k_{FN}^2 +m_N^{* 2}}$, and the Fermi momentum 
and effective mass for nucleons are $k_{FN}$ and $m^*_N = m_N - g_{\sigma N}\sigma$, 
respectively. The effective $\rho$-meson mass is 
$m_\rho^{* 2} = m_\rho^2 + 2g_{\rho N}^2(\Lambda_v g_{\omega N}^2 \omega_0^2)$. 
In asymmetric nuclear matter, the density dependence of symmetry energy
has been studied \cite{SharmaPLB,Horowitz} by simultaneously varying 
$\Lambda_v$ and $g_{\rho N}$ so that for all combinations of ($\Lambda_v$, $g_{\rho N}$) 
the symmetry energy is fixed at $E_{\rm sym}(\overline\rho,T=0) = 26$ MeV 
at an average density $\overline\rho$ corresponding to $k_{FN}=1.15$ fm$^{-1}$.
Note the symmetry energy even at the saturation density has not been
accurately constrained experimentally and only an average $E_{\rm sym}$ at $\rho_0$ 
is constrained by the binding energy of nuclei. Thus by following the above prescription 
\cite{Horowitz} one ensures accurate binding energies and proton densities of heavy 
nuclei such as $^{208}$Pb.
In fact, only those $\Lambda_v$ values were selected that reproduced the two 
measured observables mentioned above. For the present study we adopt 
$\Lambda_v = 0.0-0.03$ since the corresponding $E_{\rm sym}(\rho,T=0)$ 
and their slopes and curvatures are in reasonable agreement with that extracted from 
neutron skin thickness of several nuclei as well as the isoscaling and isospin
diffusion data \cite{SharmaPLB}. 
It may be noted that with increasing $\Lambda_v$ the density dependence of symmetry 
energy becomes {\it overall} softer, i.e., softer at supranormal densities and 
stiffer at $\rho\lesssim \rho_0$. Within this limited $\Lambda_v$ range,
considerable effect on the features of liquid-gas phase transition was also
observed in hot asymmetric nuclear matter \cite{SharmaPRC}.

The hyperon-meson coupling constants, $g_{\sigma Y}$ and $g_{\omega Y}$, are
strongly correlated by the potential depth of the hyperon $Y$ in saturated nuclear 
matter \cite{Schaffner} as
\begin{equation} \label{ypot}
U_Y^{(N)} =  g_{\omega Y}\omega_0 - g_{\sigma Y}\sigma . 
\end{equation}
We use SU(6) symmetry for the vector coupling constants
\begin{eqnarray} \label{coup}
&& \frac{1}{3} g_{\omega N} = \frac{1}{2} g_{\omega \Lambda} = 
\frac{1}{2} g_{\omega \Sigma} = g_{\omega \Xi} ~; \nonumber \\ 
&& g_{\rho N} = \frac{1}{2} g_{\rho \Sigma} = g_{\rho \Xi} ; ~~~ 
g_{\rho \Lambda} = 0 ~.
\end{eqnarray}
The $g_{\sigma Y}$ couplings are adjusted to reproduce the estimated hypernuclear 
potential depth $U^{(N)}_Y$ at $\rho_0$. From detailed studies of $\Lambda N$ 
interaction, an attractive potential depth of $U^{(N)}_\Lambda = -30$ MeV
has been estimated \cite{Millener} so that the observed bound $\Lambda$ 
hypernuclear states are reproduced. Several
$\Xi$ hypernuclear states \cite{Dover} and free $\Xi$s produced in reactions
\cite{E224,AGS} suggest an attractive potential of $U^{(N)}_\Xi = -18$ MeV. 
Though the $\Lambda$ and $\Xi$ potential depths have been well studied,
however, data on $\Sigma$ hypernuclei are scarce and ambiguous due to strong 
$\Sigma N \to \Lambda N$ decay. While previous studies of hyperon stars have used
an attractive $\Sigma$ potential (with $U^{(N)}_\Sigma = U^{(N)}_\Lambda = -30$ MeV)
that are based on the observed bound $^4_\Sigma$He hypernucleus, recent analysis 
\cite{Batty,Friedman} of $\Sigma^-$ atomic data point to a sizable repulsive nuclear 
$\Sigma$ potential of depth $U^{(N)}_\Sigma = +30$ MeV. For the present study we
adopt this repulsive depth of the $\Sigma$ potential. This ambiguity will modify 
considerably the hypernuclear composition in the star, which however, 
has smaller effects on the bulk properties of $\beta$-equilibrated star matter.

\section{Results and Discussions}

\begin{figure}[ht]
\vspace{.2cm}

\centerline{\epsfig{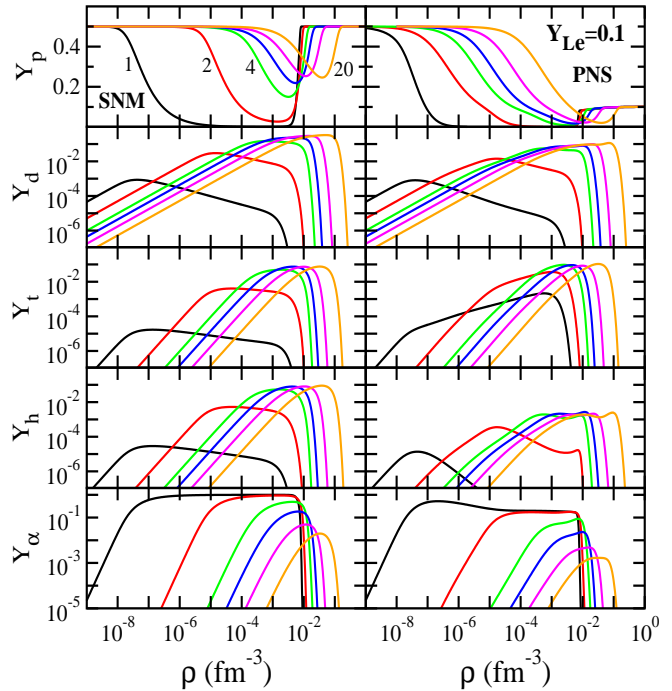}}
\caption{(Color online) Density dependence of particle fractions $Y_i$ at temperatures
$T=1,2,4,6,10,20$ MeV in symmetric nuclear matter (SNM) and in protoneutron star (PNS)
matter at a fixed lepton fraction of $Y_{Le}=0.1$ in the FSUGold model with coupling
$\Lambda_v = 0.03$.} 

\label{Yp}
\end{figure}

\begin{figure}[ht]
\vspace{.2cm}

\centerline{\epsfig{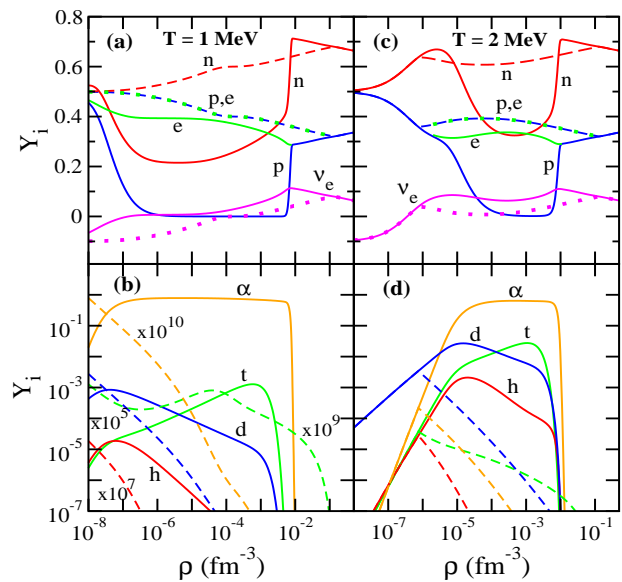}}
\caption{(Color online) Composition of supernova matter with trapped neutrinos
at a lepton fraction of $Y_{Le} = 0.4$ at a temperature of $T=1$ and 2 MeV in the 
FSUGold set for $\Lambda_v = 0.03$ coupling. The results are for uniform matter 
(solid line) and with liquid-gas phase transition (dashed line).} 

\label{Ypns}
\end{figure}

We begin this section by presenting in Fig. \ref{Yp} the fractions, 
$Y_i=A_i\rho_i/\rho$, of free protons
and clusters $i=p,d,t,h,\alpha$ at various temperatures for symmetric nuclear matter 
(SNM: left panels) and neutrino-trapped protoneutron star (PNS: right panels) matter in 
the FSUGold relativistic mean-field model for the isovector coupling $\Lambda_v = 0.03$.
The overall trend exhibited by symmetric matter is quite similar to that for
neutron-rich protoneutron star matter. At very low densities $\rho \lesssim 10^{-7}$ 
where the meson fields and the baryon chemical potentials drop rapidly 
the PNS behaves as SNM with $Y_p=0.5$ (see also Fig. \ref{Ypns}).
With the formation of the clusters, the threshold density of which is quite sensitive
to the system temperature, the free proton fraction decreases rapidly. 
At a given temperature, the two-nucleon correlation results in the formation of 
deuteron first, followed by three-particle $h,t$ and finally the $\alpha$ particle. 
At $T < 7$ MeV, the strongly bound alpha is seen to be the most abundant particle 
over a wide density range as many-body correlations gain prominence with density.
Moreover, with increasing temperature, the $\alpha$ fraction decreases while other 
cluster abundances increase; the peak positions gradually shifting to higher densities.
Consequently with increasing $T$, the minimum of $Y_p$ increases and shifts
to higher densities and then decreases again at $T \gtrsim 7$ MeV when deuterons dominate.
This was also observed in the RMF model with density-dependent couplings \cite{Typel}. 
Note that at a given temperature and density the (free) neutron-rich supernova matter 
has somewhat smaller cluster fractions compared to the symmetric nuclear matter.
The clusters eventually dissolve due to Pauli blocking at higher densities 
$\rho \geq 10^{-2}$ fm$^{-3}$ resulting in a homogeneous matter with $Y_p=0.5$ 
for SNM (left panel) and a neutron-rich $n$-$p$-$e$-$\nu$ PNS matter (right panel).

\begin{figure}[ht]
\vspace{.2cm}

\centerline{\epsfig{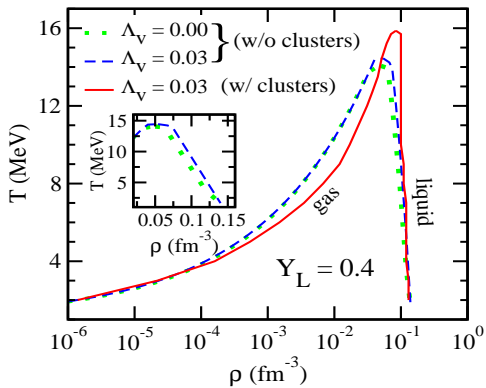}}
\caption{(Color online) Boundary of liquid-gas phase transition in a supernova matter
at a fixed lepton fraction of $Y_{Le}=0.4$ in the FSUGold model with couplings
$\Lambda_v = 0.0$ and 0.03 with and without the inclusion of clusters. The inset 
shows the symmetry energy effects for $\Lambda_v = 0.0$ and 0.03 at the liquid-phase 
boundary without clusters.} 

\label{pbound}
\end{figure}

We now consider the possibility of liquid-gas phase (LGP) transition in 
hot and neutrino-trapped protoneutron star consisting of neutrons,
protons, electrons and electron neutrino. Figure \ref{pbound} shows the boundary
for the existence of the coexistence phase at various temperatures in the FSUGold
model for the isovector coupling $\Lambda_v = 0.0, 0.03$.
The results are for electron lepton fraction $Y_{Le} \equiv Y_L = 0.4$ 
(and muon lepton number $Y_{L\mu} = 0$) which approximately corresponds to the value 
at the onset of trapping. At a nucleon density of $\rho \approx 0.08$ fm$^{-3}$, 
the critical temperature for $\Lambda_v=0.03$ is $T_c =14.6$ MeV which is close 
to that for symmetric nuclear matter \cite{SharmaPRC}. Whereas a softer nuclear 
symmetry energy $E_{\rm sym}$ at subsaturation density ($\Lambda_v=0$), gives a 
somewhat smaller $T_c$. In fact, the critical temperature divide the coexistence
boundary into a low-density gas phase (on the left) and a high-density liquid phase
(on the right) and the liquid-gas phase coexist within the boundary. Due to trapping,
the net electron concentration is quite high, and the matter at the low baryon density
phase is controlled by non-degenerate electrons. As a consequence, the nuclear symmetry
energy corresponding to $\Lambda_v=0$ and 0.03 has practically no effect on
the gas phase boundary. On the other hand, due to dominant contribution of the
degenerate nucleons in the relatively high-density liquid phase, larger effects
from symmetric pressure in the stiffer $E_{\rm sym}$ for $\Lambda_v=0.03$, shifts 
the liquid-phase boundary to a slightly higher density (see inset of Fig. \ref{pbound}). 
The protoneutron star matter is found to exhibit a LGP transition even at very low 
densities at a high temperature of $T \sim 1$ MeV.

With the inclusion of light clusters (solid line in Fig. \ref{pbound}), the softening
of the EOS shifts the low-density coexistence phase boundary to a higher value at
temperatures $T \geq 4$ MeV where clusters become abundant.
On the other hand, the high-density boundary that is dominated by free nucleons is
relatively unaffected. Thus the clusters effectively reduce the width of the phase 
coexistence region and increases the critical temperature to $T_c =15.7$ MeV 
for $\Lambda_v=0.03$ case.

In Fig. \ref{Ypns}, the particle fractions are shown for neutrino trapped matter
at an electron lepton fraction $Y_{Le} = 0.4$ at temperatures of $T=1$ and 2 MeV for
uniform matter (solid line) and in the coexistence
liquid-gas phase (dashed line). One of the major effects of trapping is to enhance the
net electron fraction $Y_e$ over the entire baryon density range; the magnitude of
$Y_e$ gradually increases with $Y_L$. Considering first uniform matter (solid line),
at $\rho \lesssim 10^{-7}$ fm$^{-3}$, the energy and pressure of the hot protoneutron 
star matter is entirely dominated 
by the non-degenerate electrons because of their small mass. Due to charge neutrality,
the rapid rise of net electrons at this low density enforces an equally large enhancement
of net number of free protons and a consequent drop in free neutron density due to
baryon number conservation (see Figs. \ref{Ypns}(a) and \ref{Ypns}(c)). This tends to make the 
lepton-rich matter in this density regime highly symmetric with identical free 
neutron and proton fractions.

For uniform matter at higher densities, clusters dominate the particle yield 
(compared to free protons and neutrons especially at high $T$; 
(Figs. \ref{Ypns}(b) and \ref{Ypns}(d)) 
so that the charge neutrality is maintained primarily by the bound protons 
in these clusters. Compared to $Y_{Le} =0.1$ (Fig. \ref{Yp}), larger electron fraction 
in the present $Y_{Le} =0.4$ case enhances the cluster production via charge neutrality.
At all temperatures of interest in the supernova dynamics, we find $\alpha$ and deuteron 
to be the dominant clusters. Thus, within this density range the star matter is also highly 
symmetric but due to abundant symmetric cluster formation.

\begin{figure}[ht]
\vspace{.2cm}

\centerline{\epsfig{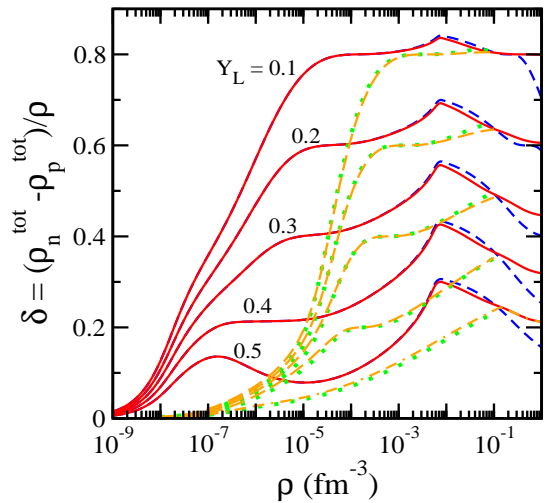}}
\caption{(Color online) Density dependence of nuclear asymmetry $\delta$ 
for supernova matter with trapped neutrinos at various lepton fraction of $Y_{Le}$
at a temperature of $T=1$ MeV. The results are in the FSUGold set at 
$\Lambda_v = (0.0, 0.03)$ for uniform matter (dashed, solid line) and with 
liquid-gas phase transition (dotted, dash-dotted line).} 

\label{asym}
\end{figure}

With the onset of liquid-gas phase transition, it is important to note from Fig. \ref{Ypns}
that cluster formation drops drastically (dashed lines). This stems from vanishing 
cluster contribution from the liquid part of the mixed phase that has a density well above 
the transition value where the clusters dissolve, as well as due to small cluster
fraction in the gas phase that has a density smaller compared to the total $\rho$.
Thus at small $T \lesssim 4$ MeV in the coexistence phase, $Y_p$ is effectively enhanced 
and becomes identical to $Y_e$. This leads to the formation of $n$-$p$-$e$-$\nu$ matter
that tends to be rather symmetric even in the coexistence phase.
However, with increasing $T$, as the cluster fractions are large and the onset of phase 
transition shifts to higher $\rho$ (see Fig. \ref{pbound}), considerable amount of clusters
can occur in the low-density uniform phase of a hot protoneutron star.
The formation of clusters and near symmetrization of the PNS matter has two major consequences. 
First, during the evolution of the supernova matter, the abundant light nuclei produced 
within the neutrino-sphere \cite{Burrows,Keil} at densities $\rho \lesssim 10^{-5}$ fm$^{-3}$ 
may be ejected outside and contribute to the nucleosynthesis of heavy elements. 
Second, the decrease in the pressure due to the nuclear 
symmetry energy overwhelms the total increase from the thermal and leptonic pressures.
This makes the equation of state softer with or without phase transition compared to the 
neutrino-free matter. 

It may be noted from 
Fig. \ref{Ypns}, that at a fixed $Y_L$, anti-neutrino $\bar\nu_e$ dominates over its
particle counterpart at lower densities and/or in the coexistence region. Moreover, 
the trapping density is also shifted to a higher value in the coexistence phase.
Thus, apart from neutrino transport, the exact location of the neutrino-sphere is 
quite sensitive to the influence of the coexistence phase in the supernova matter.

\begin{figure}[ht]
\vspace{.2cm}

\centerline{\epsfig{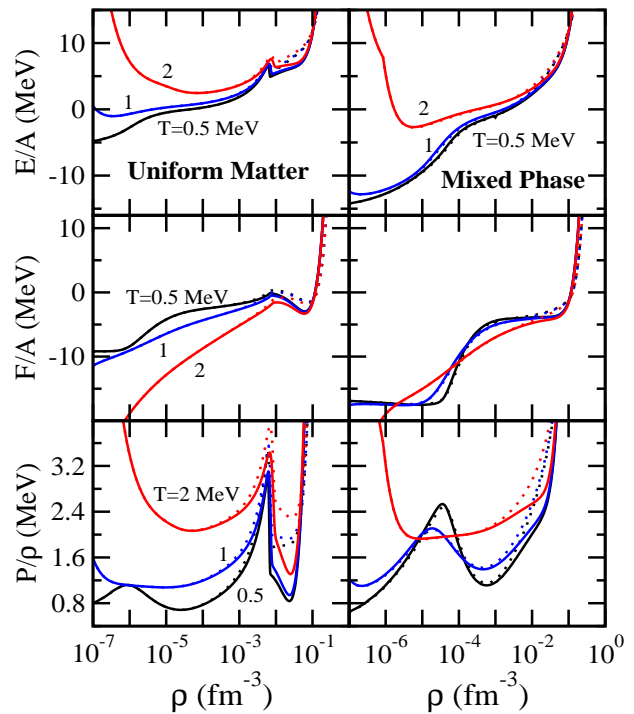}}
\caption{(Color online) Per particle energy $E/A$, free energy $F/A$, 
and pressure $P/\rho$ as a function of density for supernova matter at a 
lepton fraction $Y_{Le}=0.1$ and at temperatures $T=0.5,1,2$ MeV.
The results are in the FSUGold set at couplings $\Lambda_v = 0.0$ (solid lines)
and 0.03 (dotted lines) for uniform matter and with liquid-gas phase transition.} 

\label{SNeos}
\end{figure}

Figure. \ref{asym}, shows the variation of neutron-proton asymmetry 
$\delta = (\rho_n^{\rm tot} - \rho_p^{\rm tot})/\rho$
with baryon density at several fixed lepton fraction $Y_{Le}$ that correspond to the
deleptonization epoch of the supernova at $T=1$ MeV. Considering the 
$\Lambda_v=0.03$ set, the dip seen in the asymmetry for uniform matter (solid lines)
at $\rho \sim 10^{-2}$ fm$^{-3}$ is due to cluster formation that increases with $Y_{Le}$. 
A further rapid drop in the asymmetry at $\rho \lesssim 10^{-7}$ fm$^{-3}$ results 
from lepton dominance in the star.
Thus, even in the late stage of neutrino diffusion (at small $Y_L$), abundant nuclei 
can be produced in the near-symmetric matter \cite{Ishi,Botvina}. With the onset of 
liquid-gas phase transition at $\rho_{\rm th} \sim 0.1$ fm$^{-3}$ (dash-dotted line), 
the asymmetry is found to decrease 
faster compared to the uniform matter. This stems from dominant contribution from 
the liquid part of the mixed phase which exhibits a dropping asymmetry at 
$\rho \gtrsim \rho_{\rm th}$ (solid lines) plus a fraction from the very low-density 
region of the gas phase that is nearly-symmetric.
Both these effects systematically favor a free energy with small $\delta$.
Thus, compared to uniform matter, the supernova matter at $T \sim 1$ MeV is found to 
be nearly symmetric in the coexistence phase at larger densities near the neutrino-sphere
and the PNS behaves as $n$-$p$-$e$-$\nu$ matter. Of course at higher temperatures,
the cluster contribution would be enhanced and the onset of coexistence phase is also 
delayed to a higher density; see Figs. \ref{Ypns}(c)-(d). At a given $\rho$, this 
causes the PNS to be somewhat less symmetric compared to that at lower temperatures.

The sensitivity of $E_{\rm sym}$ on asymmetry $\delta$ has been also studied
by considering the $\Lambda_v =0$ set which gives an overall stiffer
density dependence of symmetry energy. In this case, for uniform matter (dashed line),
a larger (smaller) $\delta$ is obtained at densities smaller (larger) than the normal 
nuclear matter value. However, the effects of symmetry energy is negligible both at 
small densities that are dominated by leptons, and also in the mixed phase (dotted line).

Figure \ref{SNeos} depicts the equation of state (EOS) of supernova matter
at $Y_{Le}=0.1$ for uniform matter and in the coexistence phase for
$\Lambda_v = 0.0$ (solid lines) and 0.03 (dotted lines) with clusters.
At supranormal densities, the matter is characterized by
repulsive $\omega$-meson interaction that causes a rapid increase in
the EOS. At about the normal density of $\rho_0 \approx 0.15$ fm$^{-3}$,
the uniform matter shows a minimum for relatively small temperatures.
This is a manifestation of the saturation properties of nuclear matter when 
thermal effects contribute modestly. A larger $E_{\rm sym}$ for the 
$\Lambda_v = 0.03$ at $\rho < \rho_0$ (dotted line) gives a smaller binding
energy. In fact, the sensitivity of symmetry energy on the EOS persists till 
the contribution from free nucleons and/or clusters dominate up to a density of 
$\rho \sim 10^{-4}$ fm$^{-3}$.

\begin{figure}[ht]
\vspace{.2cm}

\centerline{\epsfig{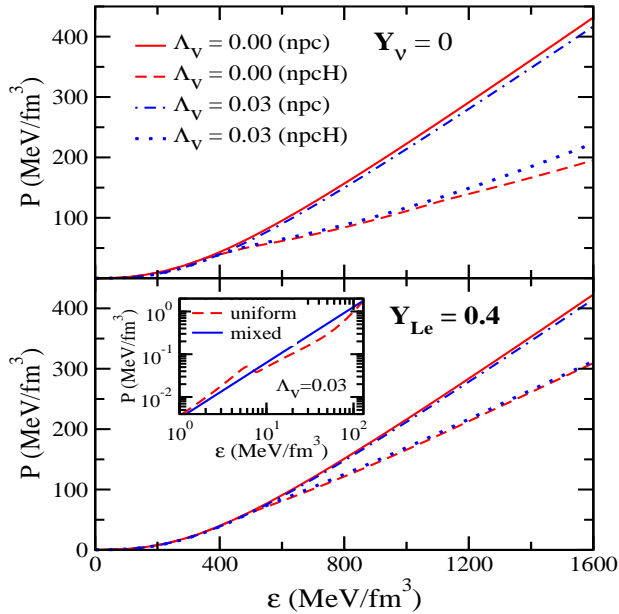}}
\caption{(Color online) Pressure versus energy density for $\beta$-equilibrated matter
with free nucleons and clusters only (npc) and with further inclusion of hyperons (npcH). 
The results are for cold lepton-poor matter ($Y_\nu=0$) and lepton-rich matter 
($Y_{Le}=0.4$) at finite entropy per baryon $S/B=1$ in the FSUGold set at $\Lambda_v = 0.0$ 
and 0.03 values for the coupling constant. The inset shows the equation of state of the 
crust for uniform matter (dashed line) and with liquid-gas mixed phase (solid line) 
at $\Lambda_v = 0.03$.}

\label{eos}
\end{figure}

In the coexistence phase, the (free) energy drops appreciably with decreasing 
density at small temperatures. As mentioned above, this is a consequence of the 
liquid part that contributes to enhance the binding of the system. 
The overall qualitative features found in the thermodynamic quantities with
clusters are similar to those without clusters for the supernova matter (not shown).
Compared to nucleons only supernova matter, the inclusion of cluster degrees of freedom 
reduces somewhat the total free energy. Further, the total pressure is found 
to be small at $\rho \sim 10^{-5}$ fm$^{-3}$ when most of the cluster species 
contribute collectively. The pressure eventually becomes stiffer when only the 
massive alpha contributes dominantly at $\rho \sim 10^{-2}$ fm$^{-3}$
(see inset of Fig. \ref{eos}). Consequently, clusters delay the onset of the 
coexistence phase (situated in the soft low $\rho$ regime of the EOS) to a higher 
density compared to that without clusters.

\begin{figure}[ht]
\vspace{.2cm}

\centerline{\epsfig{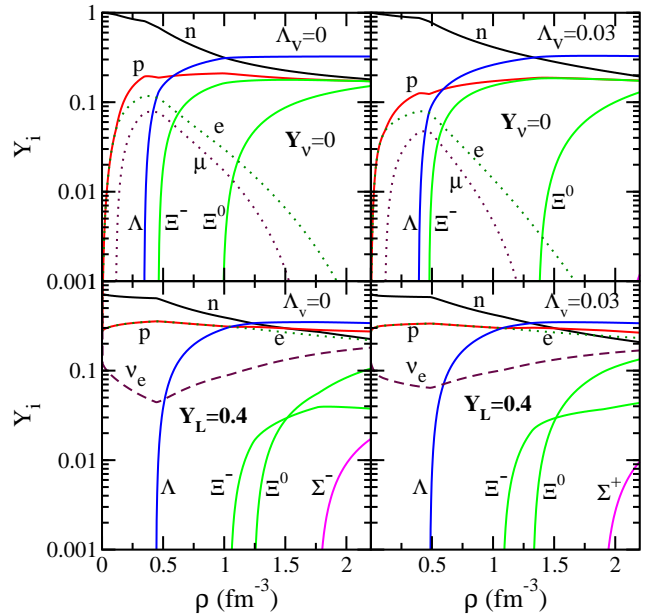}}
\caption{(Color online) Relative concentration of beta-equilibrated matter
containing nucleons and hyperons in cold lepton-poor matter ($Y_\nu=0$) and 
lepton-rich matter ($Y_{Le}=0.4$) at finite entropy per baryon $S/B=1$ in 
the FSUGold set at $\Lambda_v = 0.0$ and 0.03 values for the coupling constant.}

\label{npH}
\end{figure}

We now consider the constitution and structure of beta-equilibrated and charge
neutral matter in the superdense regime. It may be noted that thermal effects
contribute negligibly in this density regime where the nucleons become degenerate.
In Fig. \ref{npH}, the abundances of baryons and leptons as a function of density
are shown for the FSUGold parameter sets $\Lambda_v=0$ and 0.03 with and without
the trapped neutrinos. The composition of the matter should be quite sensitive
to the symmetry energy $E_{\rm sym}$ as it relates to the chemical potential
via $\hat\mu \equiv \mu_n-\mu_p \simeq 4E_{\rm sym}\delta$ \cite{Baran,BALi08} 
which in turn, determines the hyperon production threshold density by the condition
$\mu_B = \mu_n - q_B (\mu_e - \mu_{\nu_e}) \geq \varepsilon_B$, where 
$\varepsilon_B$ is the energy density of the baryon species $B$. For neutron
star matter with $Y_\nu=0$, the stiffer density dependence of symmetry energy at 
$\Lambda_v=0$ causes larger enhancement of electron (and hence proton) fraction 
compared to $\Lambda_v=0.03$ case. However, with the appearance of negatively charged $\Xi^-$
hyperon which competes with leptons in maintaining charge neutrality, the lepton
($e^-$ and $\mu^-$) concentrations begin to fall. Consequently, the sensitivity
of $E_{\rm sym}$  on the composition is diminished. The charge neutral $\Xi^0$ hyperon 
then appears, which occurs at a somewhat earlier density for the stiff symmetry energy 
compared to the soft $\Lambda_v=0.03$ set.

Trapped neutrinos have a large influence on the composition of the protoneutron star
as evident from Fig. \ref{npH} for $Y_{Le}=0.4$. The appearance of hyperons is now delayed
to higher densities since in the above mentioned threshold condition, the term 
$(\mu_e - \mu_{\nu_e})$ turns out to be positive. With the emergence of 
$\Lambda$ hyperon, the dropping of proton and thereby the electron fraction is 
now compensated by the rise of $\nu_e$ abundances so that 
$Y_{Le}$ is maintained constant. A softer symmetry energy ($\Lambda_v=0.03$) is found to
favor a significantly larger $Y_{\nu_e}$. Incidentally, at very high densities
$(\mu_e - \mu_{\nu_e})$ becomes negative. The impact of this alteration in the lepton
chemical potentials is the emergence of positively charged $\Sigma^+$ before the 
$\Sigma^{0,-}$ hyperon. From the central densities of the maximum mass stars 
(see Table I), it is evident that for the repulsive potential potential of 
$U_\Sigma^{(N)} = +30$ MeV used here, $\Sigma$ hyperons will not be present in the 
star matter at all. On the other hand, if we adopt an attractive interaction potential 
of $U^{(N)}_\Sigma = -30$ MeV, as used in previous studies, the $\Sigma^-$ hyperon 
will appear at a density of $(2-3)\rho_0$ followed by $\Sigma^0$ and $\Sigma^+$ in 
the star matter. In such a situation $\Xi$s will not appear in neutrino-trapped matter.

\begin{figure}[ht]
\vspace{.2cm}

\centerline{\epsfig{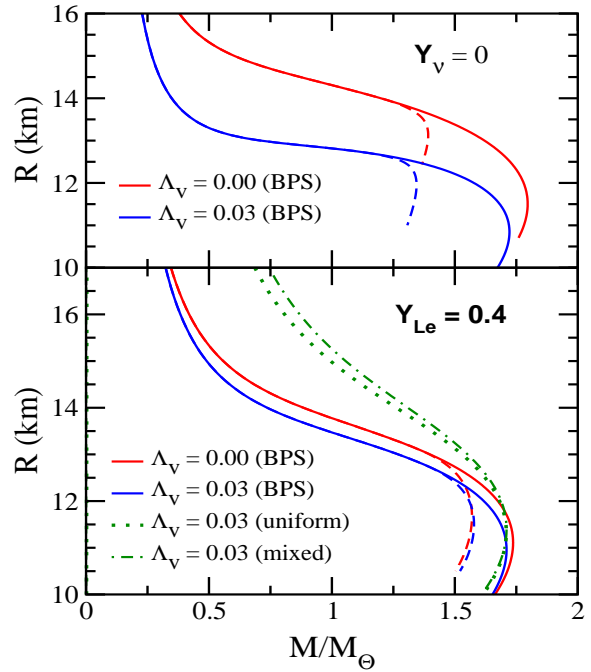}}
\caption{(Color online) The mass-radius relation for star sequences for matter
with free nucleons and clusters only (npc: solid line) and with further inclusion of 
hyperons (npcH: dashed line) in a cold lepton-poor matter ($Y_\nu=0$) and lepton-rich matter 
($Y_L=0.4$) at finite entropy per baryon $S/B=1$ in the FSUGold set at $\Lambda_v = 0.0$ 
and 0.03 values for the coupling constant with BPS equation of state \cite{BPS} at the crust. 
At $Y_L=0.4$, the results for npc star at $\Lambda_v = 0.03$ with an uniform matter (dotted line) 
and liquid-gas mixed phase (dash-dotted line) EOS at the surface are also shown; 
see text for details.}

\label{MR}
\end{figure}

The equation of state (EOS), i.e. pressure $P$ versus energy density $\varepsilon$ is
displayed in Fig. \ref{eos} for the neutrino-free matter ($Y_\nu$) and matter
with an electron lepton fraction $Y_{Le}=0.4$ corresponding to that of Fig. \ref{npH}.  
For nucleons-only star, trapped neutrinos makes the EOS softer as the decrease
in the symmetry pressure due to enhanced proton fraction exceeds the increase in
leptonic pressure. Compared to the stiff $\Lambda_v=0$ set, a smaller symmetry energy/pressure 
at high density for the $\Lambda_v=0.03$ set obviously produces a softening in the EOS. 
However, as a soft $E_{\rm sym}(\rho)$ favors a large $\nu_e$ fraction \cite{Pons},
the softening is reduced in the lepton-rich matter. With the inclusion
of hyperons, the EOS becomes softer in both lepton-rich and lepton-poor matter.
This is essentially caused by the release of the Fermi pressure of the nucleons
into the hyperonic degrees of freedom, and also due to decrease of pressure exerted
by the leptons as these are replaced by negatively charged hyperons \cite{Prakash}.
However in a protoneutron star, the trapped-neutrinos inhibit the onset
of hyperons (see Fig. \ref{npH}) which makes the EOS with hyperons stiffer compared 
to the neutrino-free matter $-$ a reversal of behavior relative to the EOS 
for nucleons-only star matter. As evident, the comparative stiffness increases 
for the larger $E_{\rm sym}$ ($\Lambda_v=0$) that allows relatively slightly 
larger hyperon population in the cold catalyzed neutron star.

\begin{table*}
\caption{Maximum masses, $M_{\rm max}/M_\odot$, their corresponding radii,
$R_{M_{\rm max}}$, and their central densities, $\rho_c/\rho_0$, of stars without
($Y_\nu=0$) and with ($Y_{Le}=0.2, 0.4$) trapped neutrinos; the normal nuclear matter
density is $\rho_0=0.148$ fm$^{-3}$. Results are in the FSUGold mean-field model 
at $\Lambda_v=0$ and 0.03.}
\centering
\begin{tabular}{lllll|llll} \\
\hline \hline
\multicolumn{1}{c}{\hfil} & 
\multicolumn{1}{c}{\hfil} & 
\multicolumn{1}{c}{~~~~ Without} & 
\multicolumn{1}{c}{hyperons} & 
\multicolumn{1}{c}{\hfil} & 
\multicolumn{1}{c}{With} & 
\multicolumn{1}{c}{hyperons} & 
\multicolumn{1}{c}{\hfil} & \\ \hline 
\multicolumn{1}{c}{$\Lambda_v$} & 
\multicolumn{1}{c}{~~~ \hfil} & 
\multicolumn{1}{c}{~~~ $\rho_c/\rho_0$} &
\multicolumn{1}{c}{~~~ $M_{\rm max}$} & 
\multicolumn{1}{c}{~~~ $R_{M_{\rm max}}$} & 
\multicolumn{1}{c}{~~~ $\rho_c/\rho_0$} &
\multicolumn{1}{c}{~~~ $M_{\rm max}$} & 
\multicolumn{1}{c}{~~~ $R_{M_{\rm max}}$}  \\ \hline 
\hfil &~~~~ $Y_\nu=0$      &~~~~ 7.138 &~~~ 1.796 &~~~~ 11.505 &~~~~ 4.912  &~~~~ 1.392 &~~~ 13.117 \\
~~0.00  &~~~~ $Y_{Le}=0.2$  &~~~~ 7.082 &~~~ 1.749 &~~~~ 11.322 &~~~~ 6.067  &~~~~ 1.444 &~~~ 12.083 \\
\hfil &~~~~ $Y_{Le}=0.4$  &~~~~ 7.420 &~~~ 1.737 &~~~~ 11.108 &~~~~ 6.716  &~~~~ 1.569 &~~~ 11.674 \\ 
\hline
\hfil &~~~~ $Y_\nu=0$      &~~~~ 7.815 &~~~ 1.722 &~~~~ 10.840 &~~~~ 5.814  &~~~~ 1.344 &~~~ 11.953 \\
~~0.03 &~~~~ $Y_{Le}=0.2$  &~~~~ 7.815 &~~~ 1.711 &~~~~ 10.747 &~~~~ 6.856  &~~~~ 1.459 &~~~ 11.341 \\
\hfil &~~~~ $Y_{Le}=0.4$  &~~~~ 7.618 &~~~ 1.710 &~~~~ 10.952 &~~~~ 6.858  &~~~~ 1.578 &~~~ 11.502 \\ 
\hline \hline
\end{tabular}
\end{table*}


The differences in the EOS should be reflected in the structure of the stars, namely their
masses and radii. While most of the mass originate from the dense interior of the star,
the crust has a negligible contribution of $\sim 10^{-5} M_\odot$ to the total mass.
In contrast, about $40\%$ of the radius originates from the EOS at $\rho \leq \rho_0$ and
thus should be strongly influenced by the possible liquid-gas phase transition at this
density regime.  At first, the star sequence is calculated with the low density 
equation of state taken from Baym-Pethick-Sutherland (BPS) 
of Ref. \cite{BPS}; the effects of the coexistence and pure phase with clusters 
are discussed later. Here, the overall EOS is constructed by smoothly matching 
(without any discontinuity in pressure, baryon and energy densities) 
the BPS equation of state at $\rho < 0.001$ fm$^{-3}$, followed by the mixed phase 
EOS comprising of leptons and free nucleons only for the inner crust of the star in the RMF model,
and finally with the RMF model EOS at $\rho > \rho_0$, with or without hyperons.
The static star sequence obtained by solving Tolman-Oppenheimer-Volkoff
equations \cite{TOV} are shown in Fig. \ref{MR} for the different equations of state.
Listed in Table I are the maximum masses $M_{\rm max}$, corresponding radii $R_{M_{\rm max}}$,
and the central baryonic densities $\rho_c$.  With free nucleons and leptons-only star,
the trapped-neutrinos give smaller maximum mass $M_{\rm max}$ (compared to $Y_\nu=0$ case) as
the pressure from the stiff EOS can support larger masses against gravitational collapse.
With the introduction of hyperons, the reversal of behavior observed in the EOS, i.e.
trapping leads to a stiffer EOS, is manifested in larger maximum mass protoneutron stars.
Note in general, the inclusion of hyperons leads to a relatively smaller mass stars due to
softening of the EOS. The larger radii in stars with hyperons are a consequence of weaker 
gravitational attraction from smaller masses that causes the stars to be large and diffuse.
For the neutrino-free stars, though $\Lambda_v=0.03$  gives smaller masses than $\Lambda_v=0$, 
the corresponding star radii are however found to be smaller in the former case.

The baryonic interactions in this model raises the maximum mass from 
$M_{\rm max} \approx 0.7 M_\odot$ for the EOS of a free gas \cite{TOV} by a factor 
of about two for nucleons-only stars. However, the inclusion of hyperons in cold 
neutrino-free stars yield maximum masses that are smaller than the current observational 
lower limit of $1.44 M_\odot$ imposed by the larger mass of the binary pulsar 
PSR $1913+16$ \cite{Weisberg}.

The inset of Fig. \ref{eos} shows the EOS for hot matter with
entropy per baryon $S/B=1$ and $Y_{Le}$=0.4 for uniform matter (dashed line) and in the 
coexistence liquid-gas mixed phase (solid line). For clarity, the EOSs corresponding
only to the high-density (liquid) regime of the mixed phase at volume fraction 
$\chi = V^G/V > 10^{-4}$ (see Eq. (\ref{free})) are shown. In fact, $\chi$ can be
as low as $10^{-8}$ near the low-density (gas) phase boundary. Figure \ref{MR} 
illustrates the effect of liquid-gas phase transition at low density on the 
mass-radius sequence of free nucleons and clusters-only protoneutron stars for 
$\Lambda_v=0.03$. Though the maximum masses are identical with 
$M_{\rm max} \approx 1.710 M_\odot$ the radii in the sequence of stars are all different. 
In particular, a stiff EOS obtained in the mixed phase from Gibbs construction gives 
consistently large radii protoneutron stars. Compared to the BPS \cite{BPS} equation of 
state used above at $\rho < 0.001$ fm$^{-3}$, the RMF models in general provide much 
stiffer EOSs at the crust that result in stars with much larger radii.

\section{Summary and Conclusion}

In this article, within an accurately calibrated relativistic mean field model
where the nuclear symmetry energy has been constrained from measurements of neutron skin
thickness of finite nuclei, the structure and properties of neutron and protoneutron
stars are studied. For hot and lepton-rich protoneutron stars we examine the
possible liquid-gas phase transition near the crust comprising of free nucleons and
light clusters. Compared to free nucleon abundances, light clusters are found to 
dominate the particle yield at moderate and high temperatures in an uniform supernova matter. 
Further, the fraction of trapped neutrino is shown to play an important role in making the 
supernova matter more symmetric via enhanced production of deuteron and alpha clusters.
With liquid-gas phase transition, the yields of clusters drop drastically at temperatures 
$T \lesssim 4$ MeV in the coexistence phase; here near symmetrization over a wide density 
$\rho \sim 10^{-5}-10^{-1}$ fm$^{-3}$ is achieved by the free nucleons that are not bound 
in the clusters. At these low densities, symmetry energy has modest impact on the coexistence 
phase boundaries and on the equations of state of matter compared to thermal effects
and on the number of trapped neutrinos.

The influence of symmetry energy on hyperon production in the dense interior
of cold neutron stars and hot lepton-rich protoneutron stars are studied. Within the 
present knowledge of the potential depth of hyperons in bulk nuclear matter, we find
a repulsive potential of $\Sigma$ hyperons inhibits its appearance in the 
star matter. Neutrino trapping delays the appearance of hyperons due to relative
disposition of chemical potentials of various species. A softer symmetry energy in general 
result in smaller chemical potentials that delays further the onset of hyperons
and thereby reduces the hyperonization of the star matter. The emergence of
hyperons soften the nuclear equation of state considerably in the FSUGold model. 
This results in stars with very small masses especially for the cold deleptonized matter.
We also find, that compared to uniform matter, a stiffer equation of state in the 
coexistence liquid-gas phase near the crust of hot lepton-rich matter yield stars
with larger radii. The ongoing and future hypernuclear programs can provide accurate 
information of the potential depth of hyperons in bulk nuclear matter, which in 
conjunction with better mass and radius measurements of neutron stars could constrain 
the nuclear symmetry energy more accurately.

\end{document}